\documentclass[prb,twocolumn,showpacs]{revtex4}
\usepackage{graphicx}
\usepackage{times}
\usepackage{dcolumn}
\usepackage{bm}

\def\be{\begin{equation}}
\def\ee{\end{equation}}
\def\bea{\begin{eqnarray}}
\def\eea{\end{eqnarray}}
\def\bse{\begin{subequations}}
\def\ese{\end{subequations}}

\def\be{\begin{eqnarray}}
\def\ee{\end{eqnarray}}

\begin{document}

\title{Goldstone modes and electromagnon fluctuations in the conical cycloid
state of a multiferroic}
\author{Sumanta Tewari$^{1,2}$}
\author{Chuanwei Zhang$^{1,3}$}
\author{John Toner$^{4}$}
\author{S. Das Sarma$^{1}$}
\affiliation{$^{1}$Condensed Matter Theory Center, Department of Physics, University of
Maryland, College Park, MD 20742\\
$^{2}$Department of Physics and Astronomy, Clemson University, Clemson, SC
29634 \\
$^{3}$Department of Physics and Astronomy, Washington State University,
Pullman, WA 99164\\
$^{4}$Department of Physics and Institute of Theoretical Science, University
of Oregon, Eugene, OR 97403}

\begin{abstract}
Using a phenomenological Ginzburg-Landau theory for the magnetic conical
cycloid state of a multiferroic, which has been recently reported in the
cubic spinel CoCr$_{2}$O$_{4}$, we discuss its low-energy fluctuation
spectrum. We identify the Goldstone modes of the conical cycloidal order,
and deduce their dispersion relations whose signature anisotropy in momentum
space 
reflects the symmetries broken by the ordered state. We discuss the soft
polarization fluctuations, the `electromagnons', associated with these
magnetic modes and make several experimental predictions which can be tested
in neutron scattering and optical experiments.
\end{abstract}

\maketitle

\section{Introduction} Although ferromagnetism and antiferromagnetism are the
two most widely studied forms of magnetic order, more complicated, spatially
modulated magnetic order parameters are also important and interesting from
both fundamental and technological perspectives.
A salient example, which occurs in the new class of `multiferroics' \cite%
{Fiebig,Ramesh,Tokura1,Cheong} -- materials that display an amazing
coexistence and interplay of long range magnetic and ferroelectric orders --
is magnetic transverse helical, or `cycloidal', order. This order
has acquired prominence \cite%
{Tokura1,Cheong,Katsura1,Mostovoy,Lawes,Tokura2,Cheong2,Chapon,Goto,Kenzelmann, Pimenov,Sneff,Tokura3,Tokura4,Dagotto,Katsura2}
since it can induce, via broken spatial inversion symmetry \cite%
{Mostovoy,Lawes}, a concomitant electric polarization ($\mathbf{P}$) in a
class of ternary oxides, leading to interesting physics of competing and
colluding ordering phenomena as well as potential applications \cite%
{Fiebig,Ramesh,Tokura1,Cheong}.
Among the exciting class of multiferroic
materials, the cubic spinel oxide CoCr$_{2}$O$_{4}$ is even more unusual,
since it displays not only the coexistence of $\mathbf{P}$ with a spatially
\textit{modulated} magnetic order, but also with a \textit{uniform}
magnetization ($\mathbf{M}$) \cite{Tokura4} in a so-called `conical cycloid'
state (see below).

Since 
in the conical cycloid state, the long range magnetic and polar orders are intertwined,
it is crucial to understand the associated soft modes (\textit{i.e.}, low
energy collective excitations), which should also be `hybridized', leading
to intriguing potential applications based on the electronic excitation of
spin waves \cite{Khitun} and vice versa.
A second motivation for studying the
soft collective mode spectrum of a system with a complicated set of
order parameters, such as the conical cycloid state, is that the
Goldstone modes themselves caricature the underlying pattern of the
broken symmetries, and thus, strengthen the understanding of the
ordered state itself.
In this paper, we do this by first identifying the magnetic Goldstone modes
(\textit{i.e.}, magnons or spin waves) of the conical cycloidal order and
deducing their dispersion relations which, as we clarify, simply reflect the
complex, anisotropic pattern of the underlying broken symmetries. We make
several predictions for inelastic neutron scattering experiments based on
our results for the magnetic fluctuations.
We then identify the associated soft polarization fluctuations, which
constitute a \textit{dielectric} manifestation of the magnetic modes,
`electromagnons', which can be observed in optical experiments. The
interesting interplay of magnons and electromagnons in cubic multiferroics
is the topic of this paper.

CoCr$_{2}$O$_{4}$, with the lattice structure of a cubic spinel, enters into
a state with a uniform magnetization at a temperature $T_{m}=93$ K.
Microscopically, the magnetization is of ferrimagnetic origin \cite{Tokura4}%
, and in what follows we will only consider the ferromagnetic component, $%
\mathbf{M}$, of the magnetization of a ferrimagnet. At a lower critical
temperature, $T_{c}=26$ K, the system develops a spacial helical 
modulation of the magnetization in a plane transverse to the large uniform
component. Such a state, for general helicoidal modulation transverse to the
uniform magnetization, can be described by an order parameter,
\begin{equation}
\mathbf{M}_{h}=m_{1}\hat{e}_{1}\cos (\mathbf{q}\cdot \mathbf{r})+m_{2}\hat{e}%
_{2}\sin (\mathbf{q}\cdot \mathbf{r})+m_{3}\hat{e}_{3},
\label{Order-Parameter-1}
\end{equation}%
where $\{\hat{e}_{i}\}$ form an orthonormal triad. When the pitch vector, $%
\mathbf{q}$, is normal to the plane of the rotating components, the rotating
components form a conventional helix \cite{Belitz}. 
A more complicated modulation arises when $\mathbf{q}$ lies \textit{in the
plane} of the rotating components. For $m_{3}=0$, we will call such a state,
which has been recently observed in a number of multiferroic ternary oxides
\cite%
{Tokura1,Cheong,Lawes,Tokura2,Cheong2,Chapon,Goto,Kenzelmann,Pimenov,Sneff,Tokura3}%
, an `ordinary cycloid' state because the profile of the magnetization
resembles the shape of a cycloid. The cycloid state with $m_{3}\neq 0$ will
be called a `conical cycloid' state, because the tip of the magnetization
falls on the edge of a cone, see Fig.~\ref{Figure:Conical}. This is the low
temperature magnetic ground state in CoCr$_{2}$O$_{4}$, and is responsible
for its many unusual properties, for e.g., the ability to tune $\mathbf{P}$
via tuning the uniform piece of the magnetization by a small magnetic field $%
\sim 0.5$ T \cite{Tokura4}. Notice that these states break the spin
rotational and the coordinate space rotational, translational and inversion
symmetries. It is easy to visualize that the helical, but not the cycloidal,
modulation preserves a residual coordinate space $U(1)$ symmetry (followed
by a translation) about the pitch vector.

\section{Intuitive understanding of the Goldstone modes} 
\begin{figure}[t]
\includegraphics[scale=0.7]{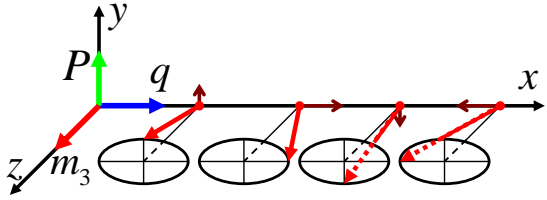}
\caption{(Color online) The conical cycloid state with the mean field order parameter given
in Eq.~\protect\ref{Order-Parameter-2}. The spins rotate along the pitch
vector, $\mathbf{q}$, on the cycloidal ($x-y$) plane. The uniform component,
$m_3$, which is along the $\hat{z}$ direction makes the magnetization tip
fall on the edge of a cone. The polarization, $P$, is perpendicular to both $%
\mathbf{q}$ and the uniform magnetization.}
\label{Figure:Conical}
\end{figure}

To gain an
intuitive understanding of the Goldstone modes, let's first consider the
broken symmetries of the conical cycloid state, with a representative
mean-field order parameter,
\begin{equation}
M_{c}(\mathbf{r})=(m_{1}\cos (qx),m_{2}\sin (qx),m_{3}),
\label{Order-Parameter-2}
\end{equation}%
shown in Fig.~\ref{Figure:Conical}. As mentioned above, this state breaks
the spin space rotation and the coordinate space rotation and translation
symmetries. Note, however, that the translation symmetry is broken only in
the direction of $\mathbf{q}$. Since translational symmetry is \textit{%
spontaneously} broken in this system, uniform translations along the
direction of ${\mathbf{q}}$, which can be parameterized by the phase
fluctuation $\varphi (\mathbf{r})$, where the fluctuating magnetization may
be given by $\mathbf{M}(\mathbf{r})=(m_{1}\cos (qx+\varphi (\mathbf{r}%
)),m_{2}\sin (qx+\varphi (\mathbf{r})),m_{3})$, must be a Goldstone mode.
It is important to realize, however, that the elastic energy for this
fluctuation \textit{cannot} involve $(\partial _{y}\varphi )^{2},(\partial
_{z}\varphi )^{2}$, while it must involve the longitudinal component, $%
(\partial _{x}\varphi )^{2}$. 
This is because a uniform rotation of $\mathbf{q}$, $\varphi (\mathbf{r}%
)=\alpha y+\beta z$, rotating the pitch vector from ($q,0,0$) to ($q,\alpha
,\beta $) must not cost any energy since the underlying Hamiltonian is
assumed to be rotationally invariant. The elastic energy must include $%
(\partial _{x}\varphi )^{2}$, however, since a change of the \textit{%
magnitude} of $\mathbf{q}$ does cost energy. Thus, in momentum space, the
dispersion relation for this Goldstone mode should be much softer in the
directions transverse to $\mathbf{q}$ than in the longitudinal direction.

The absence of a residual symmetry about $\mathbf{q}$ gives rise to a second
Goldstone mode in the conical cycloid state. Notice that a uniform rotation
of the cycloidal plane \textit{and} the uniform magnetization about $\hat{%
\mathbf{q}}$ does not cost energy, and therefore, such a rotation at long
wavelengths must cost vanishing energy. In the conventional helical state,
this mode is already contained in the phase $\varphi $, since a uniform
translation of a circular helix long its pitch axis (i.e., a uniform $%
\varphi $) is equivalent to a rotation about the pitch axis by $\varphi $.
The Goldstone mode fluctuations in the conical cycloid state are depicted
pictorially in Fig.~\ref{Figure:Softmodes}. 

\section{Ginzburg-Landau Hamiltonian} Since $\mathbf{M}$ and $\mathbf{P}$
respectively break time reversal and spatial inversion symmetry, the leading
$\mathbf{P}$-dependent piece in a Ginzburg-Landau Hamiltonian density, $%
h_{P} $, for a centrosymmetric, time reversal invariant system with cubic
symmetry is \cite{Mostovoy},
\begin{equation}
h_{P}=\mathbf{P}^{2}/2\chi +\alpha \mathbf{P}\cdot \mathbf{M}\times \nabla
\times \mathbf{M},  \label{h-P}
\end{equation}%
where $\chi >0$ and $\alpha $ are coupling constants. We assume that $%
\mathbf{P}$ is a slave of $\mathbf{M}$, in the sense that a non-zero $%
\mathbf{P}$ only occurs due to the spontaneous development of a magnetic
state with a non-zero $\mathbf{M}\times \nabla \times \mathbf{M}$. We
consider a full Hamiltonian that is \textit{completely} invariant under
simultaneous rotations of positions and magnetization. This guarantees that
any phase that can occur in our model is \textit{necessarily} allowed in a
crystal of \textit{any} symmetry. The full Hamiltonian is given by \cite{Zhang}, $H=\int
(h_{M}+h_{P})d\mathbf{r}\equiv \int hd\mathbf{r}$. Using $\mathbf{P=-}\chi
\alpha \mathbf{M}\times \nabla \times \mathbf{M}$ to eliminate $\mathbf{P}$,
we can write the total Hamiltonian density $h$ entirely in terms of $\mathbf{%
M}$,
\begin{eqnarray}
h &=&t\mathbf{M}^{2}+u\mathbf{M}^{4}+K_{0}\left( \nabla \cdot \mathbf{M}%
\right) ^{2}+K_{1}\left( \nabla \times \mathbf{M}\right) ^{2}  \nonumber \\
&&+K_{2}\mathbf{M}^{2}\left( \nabla \cdot \mathbf{M}\right) ^{2}+K_{3}\left(
\mathbf{M}\cdot \nabla \times \mathbf{M}\right) ^{2}  \nonumber \\
&&+K_{4}\left\vert \mathbf{M}\times \nabla \times \mathbf{M}\right\vert ^{2}
\nonumber \\
&&+D_{L}|\nabla \left( \nabla \cdot \mathbf{M}\right) \mathbf{|}%
^{2}+D_{T}|\nabla \left( \nabla \times \mathbf{M}\right) \mathbf{|}^{2},
\label{h-M}
\end{eqnarray}%
where we have $u$, $D_{L,T}>0$ for stability. Due to competing magnetic
interactions, some of the $K_{i}$ can be negative.

To discuss the parameter space for the conical cycloid state, $t$ is assumed
to cross zero at $T_{m}$, and the system enters into a state with a uniform
magnetization $m_{3}=\sqrt{-t/2u}$. 
As $T$ drops further,
the \textit{elliptic} conical cycloid state, with the uniform magnetization
\textit{normal} to the cycloidal plane and $\mathbf{q}$ \textit{in the plane}
of the cycloid, \textit{i.e.}, with a representative order parameter given
by Eq.~\ref{Order-Parameter-2}, is the lowest energy state in the regime $%
t<0 $, $K_{3}<0$, $K_{0}<0$, $0<K_{1}<<-K_{3}m_{3}^{2}$. In this regime,
Eq.~\ref{Order-Parameter-2} defines the ground state among all the possible
states with arbitrary mutual angles between the uniform magnetization, $%
\mathbf{q}$, and the cycloid plane. $K_{2}$, and $K_{4}$ are relatively
unimportant for this state (Eq.~\ref{Order-Parameter-2} satisfies the saddle
point equations with or without them), therefore, in what follows, we will
set $K_{2}=K_{4}=0$ for simplicity \cite{Zhang}.

\section{Goldstone modes in the conical cycloid state} To identify the
Goldstone modes and to calculate their correlation functions, we follow
standard methods: we first write $\mathbf{M}$ as its mean-field solution
(describing the conical cycloid state) plus the fluctuations. We then
substitute this total $\mathbf{M}$ in the Hamiltonian, Eq.~\ref{h-M}, and
expand the Hamiltonian to the second order in the fluctuation modes. A
straightforward (though tedious) diagonalization of the fluctuation piece of
the Hamiltonian would then produce the fluctuation modes (eigenvectors) and
their energy dispersions (eigenvalues). As we will see below, there are four
fluctuation modes of the conical cycloid state, among which two are massive
and the other two ($\alpha$ and $\delta_x$, see below) are soft (Goldstone
modes) in the long wavelength limit. By inverting the fluctuation part of
the Hamiltonian, one can also read-off the correlation functions of the soft
modes from the matrix elements.

Te begin, we write the total magnetization as $\mathbf{M}=\mathbf{M}%
_{c}+\delta \mathbf{M}$, where $\delta \mathbf{M}$ describes the
fluctuations above the saddle point solution $\mathbf{M}_{c}$. Generally, $%
\mathbf{M}$ can be written as,
\begin{equation}
M=\left(
\begin{array}{c}
-m_{3}\delta _{y}+m_{1}\cos \left( qx+\varphi \right) -m_{2}\delta _{z}\sin
\left( qx+\varphi \right) , \\
-m_{3}\delta _{x}+m_{2}\sin \left( qx+\varphi \right) +m_{1}\delta _{z}\cos
\left( qx+\varphi \right) , \\
m_{3}+\delta _{y}m_{1}\cos \left( qx+\varphi \right) +\delta _{x}m_{2}\sin
\left( qx+\varphi \right)%
\end{array}%
\right)  \label{Magnetization-Full}
\end{equation}%
where $\varphi $ describes the fluctuation of $\mathbf{q}$, and $\delta _{y}$
and $\delta _{z}$ describe the rotation of the cycloidal plane \textit{and} $%
m_{3}$ about the $y$ and the $z$ axes, respectively. Note that, for the
circular cycloidal state $(m_1=m_2)$, $\delta _{z}$ can be taken to be zero
since it only renormalizes $\varphi $ in this case.
$\delta _{x}$ describes the rotation of the cycloidal plane about the pitch
vector itself. Expanding $\mathbf{M}$ to first order in the fluctuation
variables, we have
\begin{equation}
\delta \mathbf{M}=\left(
\begin{array}{c}
-m_{3}\delta _{y}-\left( \varphi m_{1}+\delta _{z}m_{2}\right) \sin qx \\
-m_{3}\delta _{x}+\left( \varphi m_{2}+\delta _{z}m_{1}\right) \cos qx \\
\delta _{y}m_{1}\cos qx+\delta _{x}m_{2}\sin qx%
\end{array}%
\right)  \label{Magnetization-Fluctuation}
\end{equation}

To obtain the soft modes, we expand the Hamiltonian to second order in $%
\delta \mathbf{M}$. It is easy to check that the coefficient of the first
order term is zero from the saddle point equations. The second order gives%
\begin{eqnarray}
\delta H &=&t\left( \delta \mathbf{M}\right) ^{2}+u\left[ 2\mathbf{M}%
_{c}^{2}\left( \delta \mathbf{M}\right) ^{2}+4\left( \mathbf{M}_{c}\cdot
\delta \mathbf{M}\right) ^{2}\right]   \nonumber \\
&&+D_{L}|\nabla \left( \nabla \cdot \delta \mathbf{M}\right) \mathbf{|}%
^{2}+D_{T}|\nabla \left( \nabla \times \delta \mathbf{M}\right) \mathbf{|}%
^{2}  \nonumber \\
&&+K_{0}\left[ \left( \nabla \cdot \delta \mathbf{M}\right) ^{2}\right]
+K_{1}\left[ \left( \nabla \times \delta \mathbf{M}\right) ^{2}\right]
\nonumber \\
&&+K_{3}\left( \delta \mathbf{M}\cdot \nabla \times \mathbf{M}_{c}+\mathbf{M}%
_{c}\cdot \nabla \times \delta \mathbf{M}\right) ^{2}  \nonumber \\
&&+2K_{3}\left[ \mathbf{M}_{c}\cdot \nabla \times \mathbf{M}_{c}\right] %
\left[ \delta \mathbf{M}\cdot \nabla \times \delta \mathbf{M}\right]
\label{Hamiltonian-Fluctuation}
\end{eqnarray}

Substituting Eq.~\ref{Magnetization-Fluctuation} into Eq.~\ref%
{Hamiltonian-Fluctuation}, taking the Fourier transform, and denoting $%
\delta _{0}=\varphi $, we find, $\delta H=\sum_{\mathbf{p},i,j}\delta _{i}(-%
\mathbf{p})\Gamma _{ij}(\mathbf{p})\delta _{j}(\mathbf{p})$, where, $i$ and $%
j$ run from 0 to 3. For brevity, we omit the full form of the $4\times 4$
matrix $\Gamma $ here. 
\begin{figure}[t]
\includegraphics[scale=0.55]{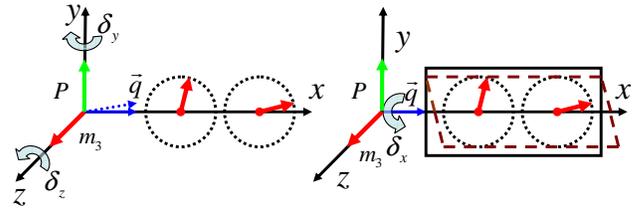}
\caption{(Color online) Left: The magnetic Goldstone mode $\protect\alpha $ of the conical
cycloid state. For an arbitrary small fluctuation of the pitch vector, $%
\mathbf{q}$, the cycloidal plane and the uniform magnetization must also
rotate by angles $\protect\delta _{y}$ and $\protect\delta _{z}$ about the
axes $y$ and $z$, respectively, for a zero energy deformation. Right: The
other Goldstone mode $\protect\delta _{x}$. $\protect\delta _{x}$ describes
the rotation fluctuation of the entire system about the pitch vector, and,
if spatially uniform, costs no energy. For clarity, the rotations of $m_{3}$
and $\mathbf{P}$ are not shown.}
\label{Figure:Softmodes}
\end{figure}

We should note, at this point, that in order for the fluctuation mode $%
\varphi $, and, in effect, the direction fluctuation of $\mathbf{q}$ to cost
vanishing energy for infinite wavelengths, the cycloidal plane and the
uniform magnetization themselves must rotate about the $y$ and the $z$ axes.
The true Goldstone mode, for the third rotation fluctuation $\delta _{x}=0$,
must then be a linear combination of $\varphi ,\delta _{y}$ and $\delta _{z}$%
. 
To capture this soft mode, we first take $\delta _{x}=0$ and diagonalize the
resulting $3\times 3$ matrix. The eigenvalues for two eigenvectors, $\beta
,\gamma $, remain non-zero even when the momentum $p\rightarrow 0$ (massive
modes), but the other eigenvalue becomes zero in this limit (soft mode). The
corresponding eigenstate of the soft mode, to linear order in $p=|\mathbf{p}%
| $, is given by,
\begin{equation}
\alpha =\varphi \left( \mathbf{p}\right) +ip_{z}\delta _{y}\left( \mathbf{p}%
\right) /q+ip_{y}\delta _{z}\left( \mathbf{p}\right) /q.  \label{Eigenvector}
\end{equation}%
This is one of the two cycloidal Goldstone modes found in this paper, see
Fig.~\ref{Figure:Softmodes}. To order $p^{2}$, we have the corresponding
eigenvalue,
\begin{equation}
\omega _{0}=2\left( m_{1}^{2}+m_{2}^{2}\right) q^{2}p_{x}^{2}.
\label{Soft-Mode1}
\end{equation}%
As expected, there is no contribution from $p_{y},p_{z}$ at this order. As
emphasized before, this is a reflection of the rotational symmetry of the
underlying Hamiltonian. The next higher order contribution to the Goldstone
mode eigenvalue is given by $\omega
_{1}=u_{1}p_{y}^{4}+u_{2}p_{z}^{4}+u_{3}p_{x}^{4}+u_{4}p_{x}^{2}p_{y}^{2}+u_{5}p_{y}^{2}p_{z}^{2}+u_{6}p_{x}^{2}p_{z}^{2}
$, where the $u_{i}$'s are functions of $m_{1},$ $m_{2}$, $m_{3}$ and the
coupling constants $K_{0}$, $K_{1}$, $K_{3},D_{L,T}$.

The other Goldstone mode of the conical cycloid state is simply the mode $%
\delta_x$, see Fig.~\ref{Figure:Softmodes}, with the momentum space
dispersion relation starting at the order $p_x^2, p_y^2, p_z^2$. As
explained before, spatially uniform rotation of the whole system about the
direction $\hat{\mathbf{q}}=\hat{x}$ does not cost energy, so the long
wavelength fluctuations, represented by $\delta_x$, cost vanishing energy.

In the presence of lattice and spin anisotropies, the foregoing results are
valid only above the anisotropy energies. The anisotropic dispersion of the
mode $\alpha$ crosses over to a more isotropic dispersion, one which depends
quadratically on all of $p_x, p_y, p_z$, below the lattice anisotropy
energy. However, it continues to remain a true Goldstone mode because of the
broken translational symmetry. In this respect, this cycloidal magnon is
analogous to the phonon mode in a crystal, rather than a true magnon mode.
Below the weak spin anisotropy energy, the other Goldstone mode, $\delta_x$,
should acquire a gap given by this spin
anisotropy energy.

In the most general case, the two soft modes will couple. In terms of the
corresponding eigenstates, the $4\times 4$ matrix can be rewritten as a $%
2\times 2$ matrix (plus unimportant contributions coming from the massive
modes),
\begin{equation}
\left(
\begin{array}{cc}
2\left( m_{1}^{2}+m_{2}^{2}\right) q^{2}p_{x}^{2}+\omega _{1} &
-p_{x}p_{z}v_{0}-ip_{x}p_{y}p_{z}v_{1}/q \\
-p_{x}p_{z}v_{0}+ip_{x}p_{y}p_{z}v_{1}/q & f(p^{2})+m_{2}^{2}D_{T}p^{4}/2%
\end{array}%
\right)  \label{Matrix-Two-Two}
\end{equation}%
where 
$v_{0},v_{1}$ are constants and $f(p^{2})$ is a second order polynomial
function of $p_{x},p_{y},p_{z}$. By inverting this matrix, we find,
\begin{eqnarray}
C_{\alpha \alpha }\left( \mathbf{p}\right) &=&\sum\nolimits_{i=x,y,z}\eta
_{i}p_{i}^{2}/\Delta \left( \mathbf{p}\right)  \nonumber \\
C_{\delta _{x}\delta _{x}}\left( \mathbf{p}\right) &=&\left[ 4\left(
g_{1}+g_{2}\right) q^{2}p_{x}^{2}+\omega _{1}\right] /\Delta \left( \mathbf{p%
}\right)  \nonumber \\
C_{\alpha \delta _{x}}\left( \mathbf{p}\right) &=&p_{x}p_{z}v_{0}/\Delta
\left( \mathbf{p}\right)
\end{eqnarray}%
where $C_{\mu \nu }\left( \mathbf{p}\right) =\left\langle \mu \left( -%
\mathbf{p}\right) \nu \left( \mathbf{p}\right) \right\rangle $, $\Delta
\left( \mathbf{p}\right) =p_{x}^{2}\sum_{i=x,y,z}\beta _{i}p_{i}^{2}+\omega
_{1}\sum_{i=x,y,z}\eta _{i}p_{i}^{2}+...$ is the determinant of the matrix (%
\ref{Matrix-Two-Two}), and the $\eta _{i}$'s and the $\beta _{i}$'s are
constants. Remarkably, for $p_{x}=0$, we find,
\begin{eqnarray}
C_{\alpha \alpha }\left( \mathbf{p}\right) &=&\omega _{1}^{-1}\sim \left(
\sum\nolimits_{i,j=y,z}p_{i}^{2}p_{j}^{2}\right) ^{-1}  \nonumber \\
C_{\delta _{x}\delta _{x}}\left( \mathbf{p}\right) &\sim &\left( \beta
_{y}p_{y}^{2}+\beta _{z}p_{z}^{2}\right) ^{-1}\sim \left(
\sum\nolimits_{i=y,z}p_{i}^{2}\right) ^{-1}  \nonumber \\
C_{\alpha \delta _{x}}\left( \mathbf{p}\right) &=&0,
\end{eqnarray}%
so there is no contribution from $p_{y}$ and $p_{z}$ to order $p^{2}$ in the
$C_{\alpha \alpha }\left( \mathbf{p}\right) $ correlator, as expected.

\section{Magnetization correlations and neutron scattering} From the energy
resolved neutron scattering cross sections near $\mathbf{p}=\mathbf{q}$, it
should be possible to track the $\mathbf{p}$-space dispersions of the
fluctuation modes $\alpha ,\beta ,\gamma $ and $\delta _{x}$ \cite{Sneff}.
Most notably, the anisotropic dispersion $\sim \omega _{0}+\omega _{1}$ of
the mode $\alpha $, caricaturing the complex broken symmetries of the
conical cycloidal order, should be experimentally testable.%

Using the soft mode eigenvectors, we can calculate the full static magnetic
susceptibility tensor, $\chi _{ij}=\left\langle M_{i}\left( -\mathbf{p}%
\right) M_{j}\left( \mathbf{p}\right) \right\rangle =\left\langle \delta
M_{i}\left( -\mathbf{p}\right) \delta M_{j}\left( \mathbf{p}\right)
\right\rangle $. For instance, the dominant terms of $\chi _{ii}$ are
\begin{eqnarray}
\chi _{xx} &\sim &m_{3}^{2}\frac{p_{z}^{2}}{q^{2}}C_{\alpha \alpha }\left(
\mathbf{p}\right) +\frac{1}{4}m_{1}^{2}\left( C_{\alpha \alpha }\left(
\mathbf{p-q}\right) +C_{\alpha \alpha }\left( \mathbf{p+q}\right) \right)
\nonumber \\
\chi _{yy} &\sim &m_{3}^{2}C_{\delta _{x}\delta _{x}}\left( \mathbf{p}%
\right) +\frac{1}{4}m_{2}^{2}\left( C_{\alpha \alpha }\left( \mathbf{p-q}%
\right) +C_{\alpha \alpha }\left( \mathbf{p+q}\right) \right)  \nonumber \\
\chi _{zz} &\sim &\frac{m_{1}^{2}p_{z}^{2}}{4q^{2}}\left( C_{\alpha
\alpha}\left( \mathbf{p-q}\right) +C_{\alpha \alpha }\left( \mathbf{p+q}%
\right) \right)  \nonumber \\
&&+\frac{1}{4}m_{2}^{2}\left( C_{\delta _{x}\delta_x}\left( \mathbf{p-q}%
\right) +C_{\delta _{x}\delta_x}\left( \mathbf{p+q}\right) \right)  \nonumber
\\
&&-\frac{p_{z}}{4q}m_{1}m_{2}\left(
\begin{array}{c}
C_{\alpha \delta _{x}}\left( \mathbf{p+q}\right) -C_{\alpha \delta
_{x}}\left( \mathbf{p-q}\right) \\
+C_{\delta _{x}\alpha }\left( \mathbf{p+q}\right) -C_{\delta _{x}\alpha
}\left( \mathbf{p-q}\right)%
\end{array}%
\right)
\end{eqnarray}%
It follows that the susceptibility functions diverge both at $\mathbf{p}=0$
and $\mathbf{p}=\pm \mathbf{q}$ for the conical cycloid state,
the divergence at $\mathbf{p}=0$ originating from the fluctuations of $m_3$.

The susceptibility functions show different behaviors when $\mathbf{p}$
approaches $\pm \mathbf{q}$ or 0 along different directions in momentum
space. For instance, when $\mathbf{p}\rightarrow \mathbf{q}$ along $p_{x}$,
all $\chi _{ii}$ diverge as $\left( p_{x}-q\right) ^{-2}$. On the other
hand, when $\mathbf{p}\rightarrow \mathbf{q}$ along $y$ or $z$ directions, $%
\chi _{xx}$ and $\chi _{yy}$ scale as $p_{i}^{-4}$, and $\chi _{zz}$ scales
as $p_{i}^{-2}$.
In neutron scattering experiments, the following quantity is related to the
frequency integrated scattering cross section \cite{Squires},
\begin{equation}
\chi \left( \mathbf{p}\right) \sim \int_{-\infty }^{\infty }d\omega \frac{1}{%
\omega }\left( 1-\exp \left( -\frac{\omega }{T}\right) \right) \frac{%
d^{2}\sigma }{d\Omega d\omega },
\end{equation}%
where $\chi \left( \mathbf{p}\right) =\left( \delta _{ij}-\frac{p_{i}p_{j}}{%
\left\vert p\right\vert ^{2}}\right) \chi _{ij}$, $\omega $ is the frequency
and $\Omega $ is a solid angle. Near $\mathbf{p}=\mathbf{q}$, the dominant
terms in $\chi \left( \mathbf{p}\right) $ are,
\begin{equation}
\chi \left( \mathbf{p}\right) \sim \frac{1}{4}m_{2}^{2}C_{\alpha \alpha
}\left( \mathbf{p-q}\right) +\frac{1}{4}m_{2}^{2}C_{\delta _{x}\delta
_{x}}\left( \mathbf{p-q}\right) .
\end{equation}%
When $\mathbf{p}\rightarrow \mathbf{q}$ along $p_{x}$, $\chi \left( \mathbf{p%
}\right) \sim \left( p_{x}-q\right) ^{-2}$. In contrast, when $\mathbf{p}%
\rightarrow \mathbf{q}$ from the $p_{y}$ or $p_{z}$ directions, the
divergence goes as $\chi \left( \mathbf{p}\right) \sim p_{i}^{-4}$.%

\section{Polarization correlations and electromagnons} The static dielectric
susceptibility tensor, $\tilde{\chi}_{ij}$, is proportional to the
polarization correlation functions, $\tilde{\chi}_{ij}(\mathbf{p})\propto
\left\langle P_{i}(-\mathbf{p})P_{j}(\mathbf{p})\right\rangle $. They can be
straightforwardly derived by using $\mathbf{P}\sim \mathbf{M}\times \nabla
\times \mathbf{M}$ and the magnon correlation functions $C_{\mu \nu }(%
\mathbf{p})$. For brevity, we do not give here the full expressions for the
polarization correlation functions. Typically, the correlation functions
transverse to $\mathbf{P}$ diverge near $\mathbf{p}=0$ and $\mathbf{p}=%
\mathbf{q}$ due to the magnetic Goldstone modes in the conical cycloid
state. Since the underlying magnons manifest themselves in the \textit{%
dielectric} response of the system, these fluctuations are sometimes called
`electromagnon' fluctuations \cite{Pimenov, Drew}.

Since the typical optical wavelengths $\sim \mathcal{O}(100$ nm$)$ are much
longer than the lattice constants $\sim \mathcal{O}(1$A$)$, we only discuss
here the behavior near $\mathbf{p}\sim 0$. Note that the fluctuations near $\mathbf{q}$ may also be
influenced by the so-called symmetric couplings between $\mathbf{P}$ and $\mathbf{M}$ \cite{Cano}, which do not contribute
to the uniform macroscopic $\mathbf{P}$. We will ignore these effects here since they are not accessible by the experiments. 
The transverse correlator along the direction of $m_{3}$
always diverges in this limit, $\left\langle P_{z}\left( -\mathbf{p}\right)
P_{z}\left( \mathbf{p}\right) \right\rangle \sim
p_{i}^{-2};(i=x,y,z,p_{j}(\neq p_{i})=0).$ This divergence arises from the
mode $\delta _{x}$, which rotates the cycloidal plane about $\hat{x}$
yielding a fluctuation of $\mathbf{P}$ along $\hat{z}$. The other transverse
susceptibility also diverges, $\left\langle P_{x}\left( -\mathbf{p}\right)
P_{x}\left( \mathbf{p}\right) \right\rangle \sim p_{y}^{-2}$, for $%
p_{x},p_{z}=0$. This divergence arises from the Goldstone mode $\alpha $.
Note that the mode $\alpha $ includes the rotation fluctuation $\delta _{z}$%
, which induces a polarization fluctuation along $\hat{x}$. These
characteristic divergences should be observable as peaks in the appropriate
static dielectric constants, revealing the existence of the electromagnon
fluctuations in the conical cycloid state. In the conical cycloid state, but
not in the ordinary cycloid state, the polarization correlation functions
diverge also near $\mathbf{p}=\mathbf{q}$, the coefficient of
proportionality of the diverging piece being $m_{3}$, but these
electromagnon fluctuations will be difficult to see in optical experiments
because of the non-zero momentum.

\section{Conclusion}To summarize, we have identified and discussed the magnetic and polarization
fluctuation modes of the conical cycloidal order in a multiferroic. One of
our primary predictions is the unusual dispersion relations of these soft
modes, which can be experimentally tested on CoCr$_2$O$_4$, thereby
revealing the complex pattern of the broken symmetries and their associated
Goldstone modes. We also predict the divergence of the magnetization and the
polarization correlation functions; the latter reveals the hybridized soft
mode, the electromagnon.

We thank D. Drew, D. Belitz, and R.Valdes Aguilar for useful discussions.
This work is supported by the NSF, the NRI, LPS-NSA, and SWAN.


\begin{thebibliography}{99}
\bibitem{Fiebig} M. Fiebig, J. Phys. D: Appl. Phys. \textbf{38}, R123 (2005).

\bibitem{Ramesh} R. Ramesh and N.A. Spaldin, Nature Materials \textbf{6}, 21
(2007).

\bibitem{Tokura1} Y. Tokura, Science \textbf{312}, 1481 (2006).

\bibitem{Cheong} S.-W. Cheong and M. Mostovoy, Nature Materials \textbf{6},
13 (2007).

\bibitem{Katsura1} H. Katsura, N. Nagaosa, and A. V. Balatsky, Phys. Rev.
Lett. \textbf{95} 057205 (2005).

\bibitem{Mostovoy} M. Mostovoy, Phys. Rev. Lett. \textbf{96}, 067601 (2006).

\bibitem{Lawes} G. Lawes, A. B. Harris, T. Kimura, N. Rogado, R. J. Cava, A.
Aharony, O. Entin-Wohlman, T. Yildirim, M. Kenzelmann, C. Broholm, and A. P.
Ramirez, Phys. Rev. Lett. \textbf{95}, 087205 (2005).

\bibitem{Tokura2} T. Kimura, T. Goto, H. Shintani, K. Ishizaka, T. Arima and
Y. Tokura, Nature \textbf{426}, 55 (2003).

\bibitem{Cheong2} N. Hur, S. Park, P. A. Sharma, J. S. Ahn, S. Guha, and
S.W. Cheong, Nature \textbf{429}, 392 (2004).

\bibitem{Chapon} L.C. Chapon, G. R. Blake, M. J. Gutmann, S. Park, N. Hur,
P. G. Radaelli, and S.-W. Cheong, Phys. Rev. Lett. \textbf{93}, 177402
(2004).

\bibitem{Goto} T. Goto, T. Kimura, G. Lawes, A. P. Ramirez, and Y. Tokura,
Phys. Rev. Lett. \textbf{92}, 257201 (2004).

\bibitem{Kenzelmann} M. Kenzelmann, A. B. Harris, S. Jonas, C. Broholm, J.
Schefer, S. B. Kim, C. L. Zhang, S.-W. Cheong, O. P. Vajk, and J. W. Lynn,
Phys. Rev. Lett. \textbf{95}, 087206 (2005).

\bibitem{Pimenov} A. Pimenov, A. A. Mukhin, V. Yu. Ivanov, V. D. Travkin, A.
M. Balbashov, A. Loidl, Nature Phys. \textbf{2}, 97 (2006).

\bibitem{Sneff} D. Senff, P. Link, K. Hradil, A. Hiess, L. P. Regnault, Y.
Sidis, N. Aliouane, D. N. Argyriou, and M. Braden, Phys. Rev. Lett. \textbf{%
98}, 137206 (2007)

\bibitem{Tokura3} Y. Yamasaki, H. Sagayama, T. Goto, M. Matsuura, K. Hirota,
T. Arima, and Y. Tokura, Phys. Rev. Lett. \textbf{98}, 147204 (2007).

\bibitem{Tokura4} Y. Yamasaki, S. Miyasaka, Y. Kaneko, J.-P. He, T. Arima,
and Y. Tokura, Phys. Rev. Lett. \textbf{96}, 207204 (2006).

\bibitem{Dagotto} I. A. Sergienko and E. Dagotto, Phys Rev. B \textbf{73},
094434 (2006).

\bibitem{Katsura2} H. Katsura, A. V. Balatsky, and N. Nagaosa, Phys. Rev.
Lett. \textbf{98}, 027203 (2007).

\bibitem{Khitun} A. Khitun and K.L. Wang, Superlattices and Microstuctures
\textbf{38}, 184 (2005).

\bibitem{Belitz} D. Belitz,
T. R. Kirkpatrick, and A. Rosch,
Phys. Rev. B \textbf{73}, 054431 (2006).


\bibitem{Zhang} C. Zhang, S. Tewari, J. Toner, S. Das Sarma, arXiv:0710.4550.

\bibitem{Squires} G.L. Squires, \textit{Introduction to the Theory of
Thermal Neutron Scattering} (Cambridge University Press, New York 1978).

\bibitem{Drew} A. B. Sushkov, R. Valde´s Aguilar, S. Park, S-W. Cheong, and H. D. Drew, Phys. Rev. Lett. \textbf{98},
027202 (2007).

\bibitem{Cano} A. Cano and E. I. Kats, Phys. Rev. B \textbf{78}, 012104 (2008).
\end{thebibliography}
\end{document}